\documentclass[twocolumn]{aastex63}
\usepackage{natbib}
\usepackage{hyperref}
\usepackage{color}
\usepackage{comment}
\usepackage{ifthen}
\usepackage{grffile}
\usepackage[utf8]{inputenc}
\usepackage{bm}
\usepackage{listings}
\usepackage{amsmath,mathtools}
\usepackage{enumitem}

\definecolor{lbcolor}{rgb}{0.9,0.9,0.9}
\specialcomment{notes}{\begingroup\sffamily\color{red}}{\endgroup}
\includecomment{comment}

\begin{document}

\title{\texttt{SPISEA}: A \texttt{Python}-Based Simple Stellar Population Synthesis Code for Star Clusters}

\author[0000-0003-2874-1196]{Matthew W. Hosek Jr.}
\correspondingauthor{Matthew W. Hosek Jr.}
\email{mwhosek@astro.ucla.edu}
\affiliation{University of California, Los Angeles, Department of Astronomy, Los Angeles, CA 90095}

\author[0000-0001-9611-0009]{Jessica R. Lu}
\affiliation{University of California, Berkeley, Department of Astronomy, Berkeley, CA 94720}
\affiliation{Institute for Astronomy, University of Hawaii, 2680 Woodlawn Drive, Honolulu, HI 96822, USA}

\author[0000-0002-6406-1924]{Casey Y. Lam}
\affiliation{University of California, Berkeley, Department of Astronomy, Berkeley, CA 94720}

\author[0000-0002-2836-117X]{Abhimat K. Gautam}
\affiliation{University of California, Los Angeles, Department of Astronomy, Los Angeles, CA 90095}

\author[0000-0002-8130-1440]{Kelly E. Lockhart}
\affiliation{Harvard-Smithsonian Center for Astrophysics, 60 Garden Street, Cambridge, MA 02138, USA}

\author[0000-0002-6658-5908]{Dongwon Kim}
\affiliation{University of California, Berkeley, Department of Astronomy, Berkeley, CA 94720}

\author[0000-0001-5341-0765]{Siyao Jia}
\affiliation{University of California, Berkeley, Department of Astronomy, Berkeley, CA 94720}

\begin{abstract}
We present \texttt{SPISEA} (Stellar Population Interface for Stellar Evolution and Atmospheres), an open-source \texttt{Python} package that simulates simple stellar populations. The strength of \texttt{SPISEA} is its modular interface which offers the user control of 13 input properties including (but not limited to) the Initial Mass Function, stellar multiplicity, extinction law, and the metallicity-dependent stellar evolution and atmosphere model grids used. The user also has control over the Initial-Final Mass Relation in order to produce compact stellar remnants (black holes, neutron stars, and white dwarfs). We demonstrate several outputs produced by the code, including color-magnitude diagrams, HR-diagrams, luminosity functions, and mass functions. \texttt{SPISEA} is object-oriented and extensible, and we welcome contributions from the community. The code and documentation are available on \href{https://github.com/astropy/SPISEA}{GitHub} and \href{https://spisea.readthedocs.io/en/latest/}{ReadtheDocs}, respectively.
\end{abstract}


\section{Introduction} 
\label{sec:intro}
The ability to simulate stellar populations is an essential tool for interpreting observations of star clusters. 
Many star clusters can be modeled as simple stellar populations (SSPs), which are characterized by a single age and metallicity.
The need for fast SSP generation has become increasingly important as the importance of stochasticity in interpreting observed populations has been more widely recognized \citep[e.g.][]{Cervino:2013lk, Krumholz:2015dk}.
Furthermore, forward modeling analysis techniques require codes that can quickly produce SSPs within likelihood functions \citep[e.g.][]{Lu:2013wo, Hosek:2019kk}.

A range of codes that simulate stellar populations are available in the literature, each offering different advantages.
Examples such as 
\texttt{Starburst99} \citep{Leitherer:1999xy, Vazquez:2005nr}, 
\texttt{PEGASE} \citep{Fioc:1997ul, Fioc:2019ix},
\texttt{GALAXEV} \citep{2003MNRAS.344.1000B}, 
\texttt{GALICS} \citep{Hatton:2003rp}, 
the evolutionary population synthesis models from \citet{Maraston:1998ol, Maraston:2005pg, Maraston:2020fj},
\texttt{PopStar} \citep{2009MNRAS.398..451M},
\texttt{Flexible Stellar Population Synthesis} \citep[\texttt{FSPS};][]{Conroy:2009le, Conroy:2010cr}, 
\texttt{CIGALE} \citep{Noll:2009if} 
and \texttt{Stochastically Lighting Up Galaxies} \citep[\texttt{SLUG};][]{da-Silva:2012hb, Krumholz:2015dk}
can be used to model SSPs and composite stellar populations (i.e., those with multiple ages and/or metallicities).
These codes are commonly used to model unresolved galaxy populations, typically offering features such as
models for dust attenuation, calculation of the photoionization and resulting nebular emission from the interstellar medium, and
prescriptions for the chemical yields of supernovae and the resulting chemical evolution of the region. 
Other codes such as \texttt{Binary Population and Spectral Synthesis} \citep[\texttt{BPASS};][]{Eldridge:2017rc, Stanway:2018dn} and 
\texttt{SYCLIST} \citep{Georgy:2014wj} specialize in SSPs, offering advanced treatment of binary stellar evolution and stellar rotation, 
respectively. \texttt{BPASS} also offers nebular emission calculations for HII regions with their model populations \citep{Xiao:2018mr}
as well as a convenient \texttt{Python} interface \citep[\texttt{Hoki};][]{Stevance:2020dp}.

However, a limitation of these codes is that they often force the user to choose between a fixed set of options when choosing the 
``ingredients'' to construct the stellar populations, such as the initial mass function (IMF), extinction law, stellar multiplicity properties,
and/or the initial-final mass relation (IFMR). 
This can be a necessary restriction due to the complexity of the underlying calculations (e.g., calculating nebular emission via radiative transfer or modeling binary stellar evolution),
but it hinders the ability to forward model these properties in observed populations. 

To address this, we present \texttt{SPISEA} (Stellar Population Interface for Stellar Evolution and Atmospheres), an open-source \texttt{Python} package that generates resolved and unresolved SSPs. 
\texttt{SPISEA} serves as a modular interface to existing stellar evolution and stellar atmosphere grids, allowing the user 
to create star clusters with control over a range of input parameters.
These span from basic properties represented by a single input value (age, metallicity, distance, extinction, differential extinction and initial cluster mass) 
to more complex properties which are represented as code objects (IMF, extinction law, multiplicity properties, photometric filters, and IFMR).
The stellar evolution and atmosphere grids are also presented as code objects that the user can select from.
This structure provides significant flexibility as the code objects are straight forward to manipulate, 
and the user can create new sub-classes in order to implement new models and/or functionalities that 
be integrated with the rest of the code.

The usefulness of \texttt{SPISEA} has been demonstrated in several published studies:
modeling the IMF of star clusters \citep{Lu:2013wo, Hosek:2019kk}, 
measuring the extinction law in highly reddened regions \citep{Hosek:2018lr}, 
predicting black hole microlensing rates \citep{Lam:2020vl}, and 
calculating photometric transformations at high extinction and with a non-standard extinction law \citep{Krishna-Gautam:2019sj, Chen:2019dp}. 
\texttt{SPISEA} can be downloaded via GitHub\footnote{https://github.com/astropy/SPISEA} with documentation provided through ReadtheDocs\footnote{https://spisea.readthedocs.io/en/latest/}. A permanent Digital Object Identifier (DOI) has been created to document this release of the code\footnote{http://doi.org/10.5281/zenodo.3937784}.

A top-level overview of the code is presented in $\mathsection$\ref{sec:overview}. 
A description of how cluster isochrones and populations are generated is provided in $\mathsection$\ref{sec:iso} and $\mathsection$\ref{sec:cluster}, respectively, and code examples are shown in $\mathsection$\ref{sec:examples}. 
In $\mathsection$\ref{sec:limit} we discuss the current limitations of \texttt{SPISEA} along with future directions for development, in $\mathsection$\ref{sec:contributions} we discuss how to contribute to the code, and then in $\mathsection$\ref{sec:conclusions} we present our conclusions.

\section{SPISEA Overview}
\label{sec:overview}
\texttt{SPISEA} has the following capabilities:

\begin{itemize}
 \item Build a theoretical cluster isochrone at a given age, distance, extinction, and metallicity. Each star is assigned intrinsic properties using metallicity-dependent stellar evolution and atmosphere grids. Synthetic photometry is calculated for a set of photometric filters defined by the user, if desired ($\mathsection$\ref{sec:iso}). 
 \item Simulate a star cluster given an isochrone at the chosen age and metallicity, initial mass, IMF, and multiplicity. Differential extinction can be added if desired. The output can either be resolved, which produces tables of physical properties and synthetic photometry for the individual star systems, or unresolved, which produces the composite spectrum of the stellar population ($\mathsection$\ref{sec:cluster}). 
 \item Calculate the population of compact stellar remnants produced by a cluster at any age using an IFMR. The type and mass of each compact object is returned ($\mathsection$\ref{sec:IFMR}). 
\end{itemize}

\begin{figure}
\begin{center}
\includegraphics[scale=0.18]{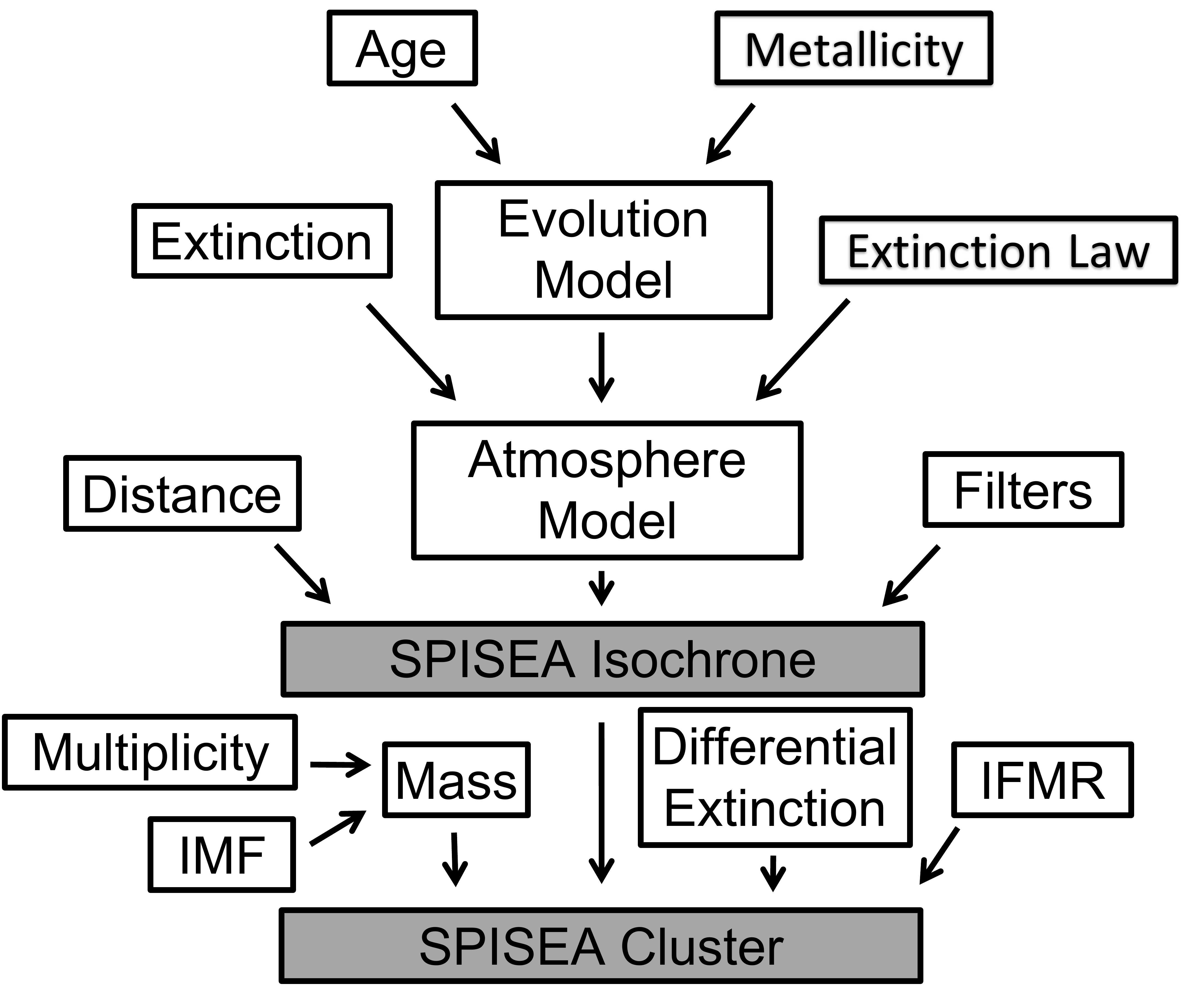}
\caption{Diagram of the \texttt{SPISEA} code. The open boxes represent inputs specified by the user. The primary outputs are the Isochrone ($\mathsection$\ref{sec:iso_output}) and Cluster ($\mathsection$\ref{sec:clust_output}) objects, which are represented by shaded boxes.
\label{fig:diagram}
}
\end{center}
\end{figure}

A top-level diagram of the workflow of the code is shown in Figure \ref{fig:diagram}. Tables with the pre-loaded set of stellar evolution and atmosphere models, extinction laws, and photometric filters are provided in Appendix \ref{app:models}.

\section{Building a Theoretical Cluster Isochrone}
\label{sec:iso}
\texttt{SPISEA} builds a theoretical cluster isochrone using existing stellar evolution and stellar atmosphere model grids.
The stellar evolution model provides basic stellar properties (e.g. current stellar mass, effective temperature T$_{eff}$, surface gravity log $g$, and luminosity $L$)
at a given age as a function of initial stellar mass. 
The stellar atmosphere model then uses this information to assign a spectrum to each stellar mass.
The primary inputs provided by the user are the isochrone age, distance, extinction, and metallicity, as well as what 
evolution and atmosphere model grids to use.
Table \ref{tab:models} describes the range of ages and metallicities available for the different evolution model grids 
in addition to the range of T$_{eff}$, log $g$, and metallicies available for the different atmosphere model grids.

\subsection{Stellar Evolution Models}
\label{sec:evo}
Stellar evolution models are defined as sub-classes off the main \texttt{evolution.StellarEvolution} object.
The user selects which model grid to use by selecting the appropriate sub-class. 
Several popular evolution model grids such as MIST \citep{Choi:2016en}, Geneva \citep{Ekstrom:2012qm}, 
and Parsec \citep{Bressan:2012ya} come pre-packaged with \texttt{SPISEA} and already have sub-classes defined (Appendix \ref{app:models}).
In addition, there are several hybrid grids that combine models across different regions of parameter space to take advantage of their individual strengths (e.g. old vs. young populations, pre-main sequence vs. main sequence stars, etc.).
These hybrid grids are discussed in Appendix \ref{app:merged}.
The user can also implement a new evolution model grid by defining their own \texttt{evolution.StellarEvolution} sub-class 
and pointing it to a directory containing the grid of isochrones produced by that model.

\subsection{Stellar Atmosphere Models}
\label{sec:atmos}
Stellar atmosphere model grids are accessed via the \texttt{get\_atmosphere} functions defined in \texttt{atmospheres.py}.
Each atmosphere model has its own \texttt{get\_atmosphere} function, with grids such as ATLAS9 \citep{Castelli:2004yq}, PHOENIX \citep{Husser:2013ts}, and BTSettl \citep{Allard:2012dq, Allard:2012zl, Baraffe:2015yg} already defined (Appendix \ref{app:models}).
Similar to the evolution models, a hybrid atmosphere model function has been defined which uses different atmosphere grids depending on the effective temperature range 
requested (Appendix \ref{app:merged}). 
In addition, the user can define their own \texttt{get\_atmosphere} functions to create a mix of the provided model atmosphere grids or implement
a new grid entirely. 

Stellar spectra are assigned from the atmosphere grid to each star in the evolution model using Space Telescope Science Institute's \texttt{pysynphot} framework \citep{STScI-Development-Team:2013fd}.
For each star, \texttt{pysynphot} searches the grid to find the best-matching atmosphere in T$_{eff}$, log $g$, and metallicity ([Z]), interpolating between models where necessary.
The spectra are originally in surface flux units (ergs cm$^{-2}$ s$^{-1}$ \AA$^{-1}$), and are multiplied by a factor of (R / D)$^2$, where R is the stellar radius (taken from the evolution model) and D is the given distance (specified by the user) to produce the \emph{unreddened} flux of the star at Earth.

The stellar atmosphere models have default wavelength range of 0.3 $\mu$m -- 5.2 $\mu$m, though the user can extend this to up to 0.1 $\mu$m -- 10 $\mu$m if desired. 
However, before extending the wavelength range, the user should confirm that their chosen atmosphere models cover the desired range (see Table \ref{tab:models}).
By default, the spectral resolution of the atmospheres have been reduced to match the ATLAS9 grid \citep{Castelli:2004yq}, which corresponds to R $\sim$ 250. 
This is generally sufficient for the purposes of synthetic photometry.
Versions of the atmosphere models at their original resolution (also in Table \ref{tab:models}) are available for download with \texttt{SPISEA}, but calculating synthetic photometry at these high resolutions is significantly slower.

\subsection{Extinction}
Extinction is applied to the synthetic stellar spectra using the total extinction and an extinction law.
The total extinction is parameterized as A$_{K}$, the total magnitudes of extinction at K-band\footnote{Note that different extinction laws use different K-band filters, and thus have different central wavelengths $\lambda_0$ such that A$_{\lambda_0}$~/~A$_{K}$~=~1. Definitions of $\lambda_0$ are provided in Table \ref{tab:extinction}.}, and the extinction law is defined as A$_{\lambda}$~/~A$_{K}$.
The total extinction at a given wavelength is thus:

\begin{equation}
A_{\lambda} = A_{K} * (A_{\lambda} / A_{K})
\end{equation}
and the reddened flux F$_r$ at $\lambda$ is:
\begin{equation}
F_r(\lambda) = F_i * 10^{-0.4 * A_{\lambda}}
\end{equation}
where F$_{i}$ is the unreddened flux of the star.

\texttt{SPISEA} again uses the \texttt{pysynphot} framework for this calculation, and the extinction law is defined as sub-class of the \texttt{pysynphot.reddening.CustomRedLaw} object. 
The set of pre-defined extinction laws include those from \citet{Cardelli:1989qf}, \citet{Nishiyama:2009fc}, and \citet{Schlafly:2016cr}. 
The user can also define a power-law extinction law with an arbitrary exponent using the \texttt{RedLawPowerLaw} sub-class.

\subsection{Synthetic Photometry}
The user can calculate synthetic photometry for the stars in the isochrone object. 
Filter transmission functions are defined as \texttt{pysynphot.ArrayBandpass} objects, which are convolved with the source spectrum to calculate the flux in a given filter. 
In this first release of \texttt{SPISEA}, the source spectra are defined between 0.25 $\mu$m -- 5.2 $\mu$m.

Stellar magnitudes are calculated in the Vega system:
\begin{equation}
m_s = -2.5 * log(F_s / F_{Vega}) + M_{Vega}
\end{equation}
where $F_s$ is the integrated flux of the source star in the filter, $F_{Vega}$ are the integrated flux of the Vega star model in the filter, and $M_{Vega}$ = 0.03 mag is the magnitude of Vega in the filter.
We adopt a Kurucz atmosphere with T$_{eff}$ = 9550 K, log $g$ = 3.95, and [Z] = -0.5 as a model for Vega \citep{Castelli:1994mi}, and assume $(R / d)^2$ = 6.247x10$^{-17}$, where $R$ is the stellar radius and $d$ is the distance of Vega \citep{Girardi:2002rb}.

Additional photometric calibrations (such as AB magnitudes) are not yet supported by \texttt{SPISEA}. If this functionality is desired, the user can request it via the Github issues page ($\mathsection$\ref{sec:contributions}).

\subsection{Isochrone Output}
\label{sec:iso_output}
A \texttt{SPISEA} isochrone is defined as a \texttt{synthetic.Isochrone} object.
If synthetic photometry is desired, then the \texttt{synthetic.IsochronePhot} sub-class should be used.
All \texttt{synthetic.Isochrone} objects contain an array with the reddened spectra for each star, while the \texttt{synthetic.IsochronePhot} sub-class contains an additional \texttt{Astropy} table \citep{Astropy-Collaboration:2013kx, Astropy-Collaboration:2018ws} with the stellar parameters and synthetic photometry in a specified set of filters.
A description of the columns in the output table is provided in Table \ref{tab:iso_tab}.

The table from \texttt{synthetic.IsochronePhot} is saved in \texttt{FITS} table format with a standard file name in a directory specified by the user.
When \texttt{synthetic.IsochronePhot} is called, it will first search the specified directory to see if the file already exists.
If it does, it will simply read the file rather than redoing the full isochrone calculation.
Thus, generating a grid of \texttt{synthetic.IsochronePhot} isochrones can save considerable time when analyzing an observed stellar population.

\begin{deluxetable}{l l c} 
\tablecaption{Isochrone Table Output}
\tabletypesize{\scriptsize}
\tablehead{
\colhead{Column} & \colhead{Description} & \colhead{Units}
}
\startdata 
L & Luminosity & W \\
Teff & Effective Temperature & K \\
R & Radius & m \\
mass & Initial Mass & M$_{\odot}$ \\
logg & Surface Gravity & cgs \\
isWR & Is star a Wolf-Rayet? & boolean \\
mass\_current & Current Mass & M$_{\odot}$ \\
phase & Evolution Phase\tablenotemark{a} & \\
m\_* & Magnitude in filters & Vega Mag \\
\enddata
\tablenotetext{a}{The returned phases are as defined by the published evolution model for all but the compact objects. For compact objects, the phases are always: 101 = white dwarf, 102 = neutron star, 103 = black hole.}
\label{tab:iso_tab}
\end{deluxetable} 

\section{Making A cluster}
\label{sec:cluster}
Once an isochrone has been created, the user can create a synthetic cluster by specifying the stellar multiplicity, IMF, IFMR, differential extinction, and initial cluster mass. 
Various aspects of generating a synthetic cluster have been described in sections of \citet{Lu:2013wo}, \citet{Hosek:2019kk}, and \citet{Lam:2020vl}, but we collate and summarize the process here.

\subsection{Multiplicity}
\label{sec:Mult}
Surveys of nearby stellar populations reveal that the fraction of multiple systems is high, rising from roughly 20\% for M $\sim$ 0.1 M$_{\odot}$ stars to nearly 100\% for M $\gtrsim$ 5 M$_{\odot}$ stars \citep[e.g.][]{Sana:2012ez, Duchene:2013bs}. 
Further, the properties of these systems have been shown to vary as a function of primary mass \citep{Moe:2017yg}. 
In \texttt{SPISEA}, one can construct a Multiplicity object (\texttt{multiplicity.MultiplicityUnresolved}) that defines the multiplicity fraction ($MF$), companion star frequency ($CSF$), and mass ratio ($q$) of synthetic cluster. 

Following \citet{Reipurth:1993lr}, the $MF$ describes the fraction of stars that host multiple systems:

\begin{equation}
MF = \frac{B + T + Q + ...}{S + B + T + Q + ...}
\end{equation}
where $S$ is the number of single stars, $B$ is the number of binary stars, $T$ is the number of triple stars, $Q$ is the number of quadruple systems, and so forth. 
The $CSF$ describes the expected number of companions in a given multiple system:

\begin{equation}
CSF = \frac{B + 2T + 3Q + ...}{S + B + T + Q + ...}.
\end{equation}
Finally, the mass ratio $q$ defines the ratio between the primary star mass and the companion star mass. 

For each star drawn from the IMF ($\mathsection$ \ref{sec:IMF}), its multiplicity properties are drawn from the $MF$, $CSF$, and $q$ distributions and companion stars assigned accordingly. 
All multiple systems are assumed to be unresolved, and the individuals fluxes of the stars are added together
to produce the final synthetic photometry of the system. 
Note that \texttt{SPISEA} does not yet include the orbital properties of the multiple systems, such as binary separation or eccentricity. 
In addition, the impact of multiplicity on stellar evolution is also ignored.
The implementation of orbital properties and binary stellar evolution models is left for future versions of \texttt{SPISEA} ($\mathsection$\ref{sec:limit}).

The default parameters for \\ \texttt{multiplicity.MultiplicityUnresolved} are defined in \citet{Lu:2013wo}, who define empirical functions for $MF$, $CSF$, and $q$ based on observations of young clusters ($<$10 Myr). The $MF$ and $CSF$ are defined as power laws as a function of the stellar mass:

\begin{equation}
MF(m) = A*m^{\alpha}
\end{equation}

\begin{equation}
CSF(m) = B*m^{\beta}
\end{equation}

where $A$ = 0.44, $\alpha$ = 0.51, $B$ = 0.50, $\beta$ = 0.45, and $m$ is the stellar mass in units of solar masses. 
The $MF$ is defined over a range from [0,1] and the $CSF$ is defined from [0, 3].
If a given stellar mass is large enough that the $MF$ of $CSF$ would be larger than their maximum values,
they are simply set to the maximum value itself.

Finally, the probability distribution for $q$ is defined as a single power law with no mass dependence:

\begin{equation}
P(q) = \left (  \frac{1 + \gamma}{1 - q_{lo}^{1 + \gamma}} \right ) * q^{\gamma}
\end{equation}

where $\gamma$ = -0.4 and $q_{lo}$, which represents the lowest allowed mass ratio, is 0.01.  

The user can change the values for $A$, $B$, $\alpha$, $\beta$, $\gamma$, max $CSF$, and $q_{lo}$ in \texttt{multiplicity.MultiplicityUnresolved} 
by adjusting the appropriate keyword arguments.

\subsection{Initial Mass Function}
\label{sec:IMF}
The IMF describes the initial distribution of stellar masses in a star cluster. 
While the true functional form(s) of the IMF is unknown, it is often described a log-normal or broken power-law distribution \citep[e.g.][]{Bastian:2010dp}. 
\texttt{SPISEA} defines the IMF as an \texttt{imf.IMF} object, and currently supports a broken power-law functional form as defined by the sub-class \texttt{imf.IMF\_broken\_powerlaw}.
The user has control over the number of power-law segments, the power-law slope for each segment, the break masses between segments, and stellar mass range the IMF is defined over. 
The \texttt{Multiplicity} object is an additional input for the \texttt{IMF} object that describes the multiplicity properties of the population ($\mathsection$ \ref{sec:Mult}).

Several standard broken power-law IMFs are included such as from \citet{Salpeter:1955yu}, \citet{Miller:1979sf}, \citet{2001MNRAS.322..231K}, and \citet{Weidner:2004ly}.
Functional forms other than broken power-laws may be added as additional sub-classes of the \texttt{imf.IMF} object, but this is beyond the scope of the initial code release ($\mathsection$\ref{sec:limit}).

\texttt{SPISEA} uses the algorithm described by \citet{Pflamm-Altenburg:2006ai} to stochastically draw stars from the IMF until the initial cluster mass is reached.
First, a rough estimate of the total number of stars in the cluster is made (the initial cluster mass divided by the average stellar mass in the IMF). 
Then stars are drawn from the IMF in batches equal to 10\% of the total number of stars and are assigned companions according to the input multiplicity model. 
This continues until the cumulative mass of all stars (including companions) is closest to the initial cluster mass.
This process is most similar to the \texttt{STOP\_NEAREST} stochastic sampling technique defined in \texttt{SLUG} \citep{Krumholz:2015dk}.

Note that no age information is used at this point and all sampling is done on the initial stellar mass and the initial cluster mass, not the present-day masses.
Once these initial masses are drawn, we use the stellar evolution model for the input population age and metallicity to determine the current properties of the stars, as discussed in section $\mathsection$\ref{sec:evo}.

\subsection{Initial-Final Mass Relation}
\label{sec:IFMR}
The IFMR maps a star's initial mass, also called the zero-age main sequence (ZAMS) mass, to the type and mass of the compact object it will form \citep[e.g.][]{Portinari:1998jr, Kalirai:2008fq, Sukhbold:2016pr, Raithel:2018mg}. 
It is an active area of study, in particular at high stellar masses where the IFMR is not well constrained. 
In \texttt{SPISEA}, the user can define an IFMR object (\texttt{ifmr.IFMR}) to simulate the compact objects produced by a stellar population.

The default \texttt{SPISEA} IFMR object is a combination of two IFMRs in the literature: one for white dwarfs (WD), and another for neutron stars (NS) and black holes (BH).
The WD IFMR comes from \citet{Kalirai:2008fq}, and is derived from observations of open clusters.
The NS/BH IFMR comes from \citet{Raithel:2018mg}, which is derived from the 1-D neutrino-driven supernova simulations of \citet{Sukhbold:2016pr} coupled with the distribution of observed NS and BH masses.
For a mathematical description of the default IFMR, see \citet{Lam:2020vl}.

The WD IFMR from \citet{Kalirai:2008fq} is relatively straightforward, as the final WD mass only depends on the ZAMS mass.
The NS/BH IFMR is more complicated as stellar metallicity, rotation, and core structure of the pre-supernova star have been found to play important roles in determining the type and mass of remnant formed \citep{Heger:2003, Sukhbold:2018}.
As a result, \citet{Raithel:2018mg} derive a probabilistic IFMR, where each ZAMS mass is assigned a probability of being a NS or BH.
In \texttt{SPISEA}, each star of sufficient ZAMS mass are designated as NS or BH according to these probabilities. 
It should be noted that the IFMR of \cite{Raithel:2018mg} is for single and solar metallicity stars.
However, there are models with different metallicities and a binary IFMR that will be published in the near future (T. Sukhbold, private communication) that will be implemented in future versions of \texttt{SPISEA}.
In addition, the user may define their own IFMR objects as they see fit.

All compact objects produced from the IFMR are assumed to be dark (e.g. zero luminosity) since we do not have the necessary evolution or atmosphere models to describe them. 
Although real WDs, NSs, and accreting BHs have non-zero luminosities, these objects are typically much fainter at optical/near-infrared wavelengths than the majority of the of the surrounding stellar population \citep[e.g.][]{Kalirai:2008fq}. 
Thus, this assumption has negligible impact on the cluster population photometry in most cases.
However, improved treatment of these sources is an avenue of future code development. 

The exception to this is the \texttt{MISTv1} evolution models, which produce model parameters for a subset of the WD population. 
These models include stars down to the white dwarf cooling phase until $\Gamma = 20$, where $\Gamma$ is the Coulomb coupling parameter \citep{Choi:2016en}. 
For these objects, a WD model atmosphere can be assigned \citep[e.g.][]{Koester:2010wd} and synthetic photometry computed. 
However, objects with $\Gamma > 20$, which are extremely cooled or crystallized, are not included in the \texttt{MISTv1} models.
In \texttt{SPISEA}, these objects will will be picked up by the \citet{Kalirai:2008fq} IFMR to produce the aforementioned dark WDs. 

\subsection{Differential Extinction}
\label{sec:dAKs}
Differential extinction is a phenomenon by which the stars within a cluster exhibit a distribution of extinction values rather than a constant value.
This could occur due to variations in the density of foreground gas and dust along the line-of-sight to a cluster, and has been observed observed in several Milky Way clusters \citep[e.g.][]{Burki:1975sj, Schodel:2010eq, Habibi:2013th, Hosek:2015cs, Andersen:2017aq, 2019ApJ...877...37R}.
The user can control the differential extinction through the $\sigma_{A_{K}}$ parameter.
For each star, \texttt{SPISEA} will perturb the total extinction A$_{K}$ (the magnitudes of extinction in K-band) by an amount $\delta A_{K}$ drawn from a Gaussian distribution with mean $\mu$ = 0 and standard deviation $\sigma$ = $\sigma_{A_{K}}$. 

To derive the reddening vector of each photometric filter, the Vega model atmosphere is extinguished at both A$_{K}$ and A$_{K}$ + $\sigma_{A_{K}}$ and the resulting change in magnitude $\delta_m$ is calculated. 
The reddening vector is then $\delta_m$~/~$\sigma_{A_{K}}$.
This reddening vector is then used to calculate the change in magnitude caused by the value of $\delta A_{K}$ drawn for each star. 
For multiple star systems, all stars within the system are assigned the same $\delta A_{K}$ value. 

Since the reddening vector is only calculated using a single stellar spectrum, variations in the vector as a function of stellar mass is ignored.
Calculating reddening vectors at each stellar mass significantly increases the computational time to create differentially reddened clusters, and its impact 
is quite small.
In an extreme example, such as a cluster with A$_K$ = 2 mag and $\delta A_{K}$ = 1 mag \citep[and assuming a][ extinction law]{Cardelli:1989qf}, the difference in $\delta_m$ for a T$_{eff}$ = 10,000 K and a T$_{eff}$ = 3,500 K star is $<$ 1\% in standard near-infrared filters (JHK). 
However, the effect is larger at shorter wavelengths, reaching 6-8\% in the V and R filters\footnote{This is not surprising, given that $\delta A_{K}$ = 1 mag corresponds to a large $\delta A_{\lambda}$ in these filters, in this case $\delta A_{V}$ $\sim$ 8.6 mag and $\delta A_{R}$ $\sim$ 7 mag.}.
For use cases where the mass-dependent extinction vector must be accounted for, we recommend generating multiple \texttt{synthetic.IsochronePhot} objects at different
extinctions to directly calculate the reddening vector at each stellar mass.

\subsection{Cluster Output}
\label{sec:clust_output}

\subsubsection{Resolved Clusters}
Resolved clusters are defined via the \texttt{synthetic.ResolvedCluster} object, which takes an \texttt{Isochrone} object, \texttt{IMF} object and the associated \texttt{Multiplicity} object, initial cluster mass,
and \texttt{IFMR} object as inputs. 
If a differential extinction is desired, then the \texttt{synthetic.ResolvedClusterDiffRedden} sub-class should be used, which takes $\sigma_{A_K}$ as an additional input.
\texttt{ResolvedCluster} objects take the \texttt{Isochrone} object and calculates a linear interpolation of all isochrone properties (e.g. T$_{eff}$, log $g$, $L$, and synthetic photometry) as a function of stellar mass. 
It then draws individual stellar masses from the IMF and multiplicity inputs and assigns properties to the stars using the interpolation functions.

The output that is produced depends on whether multiplicity is invoked. 
If no multiplicity is defined (e.g., no multiple systems), then \texttt{ResolvedCluster} will contain one \texttt{Astropy} table that contains the physical properties and synthetic photometry of the individual stars in the cluster. 
If multiplicity is defined, then the \texttt{ResolvedCluster} object will contain two \texttt{Astropy} tables: the first containing the physical properties of the primary star, the total mass of the system, and the composite synthetic photometry of the system, and the second with the physical properties and synthetic photometry of the individual companion stars.
A description of the table columns is provided in Table \ref{tab:clust_tab}.

It is worth noting that the population of stars returned in the \texttt{ResolvedCluster} output tables have slightly different interpretations depending on whether or not an IFMR is defined.
If no IFMR is defined, then the stars in the output tables are only those in the stellar evolution model, which typically exclude compact stellar remnants (\texttt{MISTv1} is an exception to this, containing a subset of the white dwarf population, e.g. $\mathsection$\ref{sec:IFMR}). 
Thus, tables contain stars that have not evolved into compact stellar remnants at the given population age.
Alternatively, if an IFMR is defined, then objects that have evolved into compact stellar remnants are assigned properties according to that IFMR.
These are included in the output tables.

\subsubsection{Unresolved Clusters}
The user can also produce an unresolved cluster with the \texttt{synthetic.UnresolvedCluster} object.
The \texttt{synthetic.UnresolvedCluster} object takes the same inputs as the \texttt{ResolvedCluster} object.
\texttt{UnresolvedCluster} produces a spectrum that is comprised of the spectra of all the stars in the cluster.
Spectra are assigned to each star based on the closest model in the stellar isochrone by mass. 


\begin{deluxetable}{l l c} 
\tablecaption{\texttt{ResolvedCluster} Table Output}
\tabletypesize{\scriptsize}
\tablehead{\multicolumn{3}{c}{Primary Star Table} \\
\colhead{Column} & \colhead{Description} & \colhead{Units}
}
\startdata 
mass & Initial Mass & M$_{\odot}$ \\
isMultiple & Is Multiple System? & boolean \\
systemMass & Total System Initial Mass & M$_{\odot}$\\
Teff & Effective Temperature & K \\
L & Luminosity & W \\
logg & Surface Gravity & cgs \\
isWR & Is star a Wolf-Rayet? & boolean \\
mass\_current & Current Mass & M$_{\odot}$ \\
phase & Evolution Phase\tablenotemark{a} & -- \\
m\_* & System magnitude in filters & Vega Mag \\
N\_companions & Number of Companions & -- \\
AKs\_f & Stellar Extinction & mag (in Ks)\\
\hline
\hline
\\
 & Companion Star Table & \\
 \\
\hline
system\_idx & Index of system in Primary Star Table & --\\
mass & Initial Mass & M$_{\odot}$ \\
Teff & Effective Temperature & K \\
L & Luminosity & W \\
logg & Surface Gravity & cgs \\
isWR & Is star a Wolf-Rayet? & boolean \\
mass\_current & Current Mass & M$_{\odot}$ \\
phase & Evolution Phase\tablenotemark{a} & -- \\
m\_* & System magnitude in filters & Vega Mag \\
\enddata
\tablecomments{The companion star table is only created if multiplicity is used. }
\tablecomments{For clusters with no multiplicity, the systemMass and m\_* columns contain the single-star results, and the N\_companions column is not created. The AKs\_f column is only returned if \texttt{ResolvedClusterDiffRedden} object is used.}
\tablenotetext{a}{The phases are as defined by the published evolution model for all but the compact objects. For compact objects, the phases are always: 101 = white dwarf, 102 = neutron star, 103 = black hole.}
\label{tab:clust_tab}
\end{deluxetable}

\section{Examples}
\label{sec:examples}
Here we provide example code for how to generate a theoretical isochrone and star cluster using \texttt{SPISEA},
as well as several plots demonstrating the outputs. 
In addition, the documentation contains several \texttt{jupyter} notebooks to help new users,
including a quick-start guide and code to reproduce the plots presented below.

\subsection{\texttt{SPISEA} Isochrones}
\begin{figure*}
\begin{center}
\includegraphics[scale=0.30]{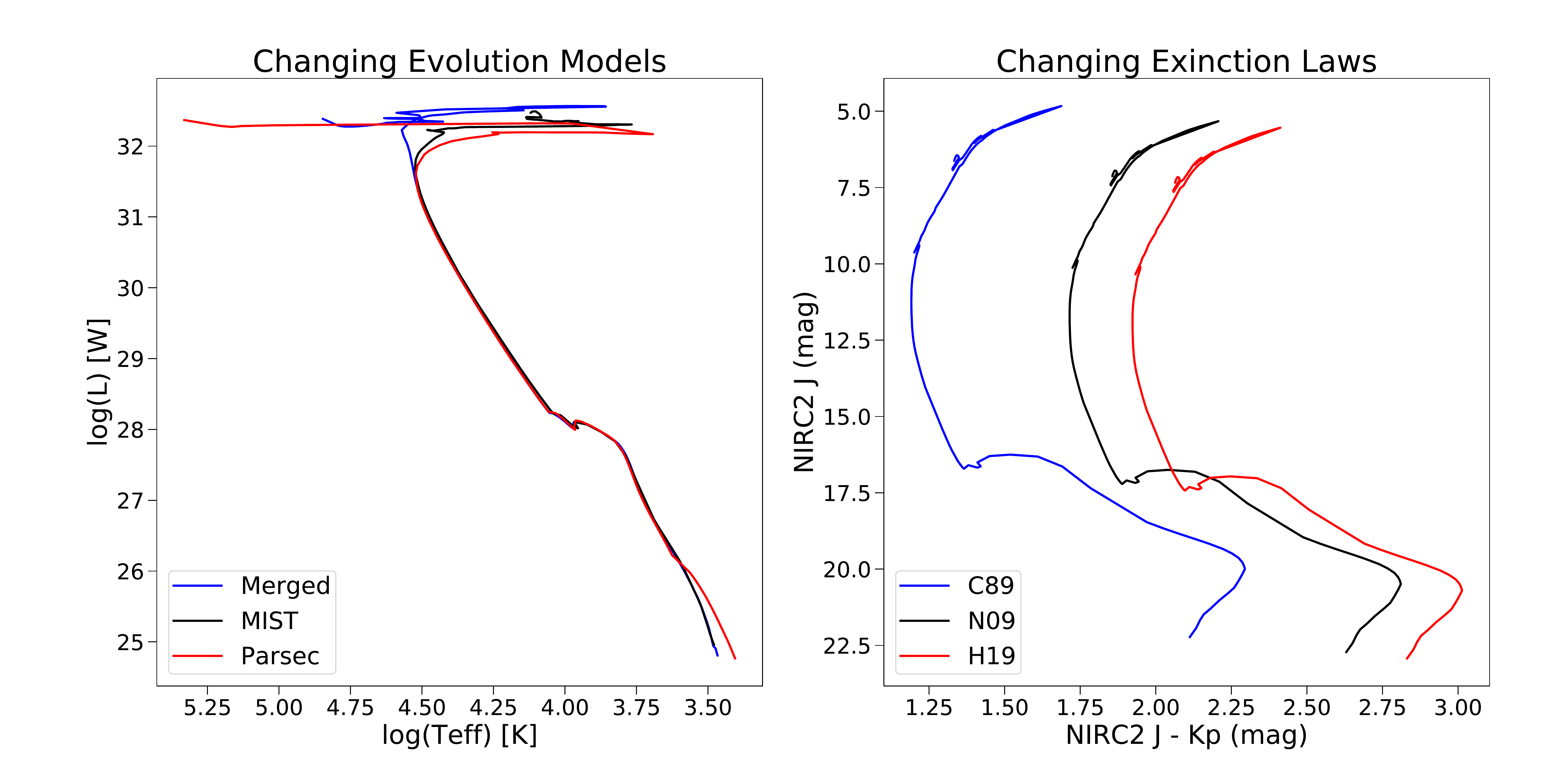}
\caption{\emph{Left:} The HR-diagram for a 5 Myr cluster isochrone at solar metallicity using the \texttt{MergedBaraffePisaEkstromPargisec} (blue), \texttt{MISTv1} (black), and \texttt{Parsec} (red) evolution models. \texttt{SPISEA} allows the user to change evolution models to examine the impact they have on cluster output. \emph{Right:} The color-magnitude diagram (Keck/NIRC2 J+Kp filters) of a cluster with an extinction of A$_{Ks}$ = 1.0 mag and distance of 4000 pc, using the \citet{Cardelli:1989qf} law (R$_v$ = 3.1; blue), \citet{Nishiyama:2009fc} law (black), and \citet{Hosek:2019kk} law (red). The choice of extinction law has a significant impact on the star colors.}
\label{fig:evo_comp}
\end{center}
\end{figure*}

The code required to produce a theoretical cluster isochrone is shown in Listing \ref{code:iso}.
As discussed in $\mathsection$\ref{sec:iso}, the user has significant control when creating an isochrone, with the ability to change the evolution models, atmosphere models, and extinction law.
These parameters are defined as \texttt{python} objects and thus can be interchanged easily. 
For example, one can change stellar evolution models to examine the impact of the different physics and assumptions used in those models (e.g. Figure \ref{fig:evo_comp}). 
This flexibility aids the assessment of systematic uncertainties introduced by these different models when interpreting observations.

Figure \ref{fig:evo_comp} also shows how changing the extinction law at a constant total extinction impacts the isochrone. 
While the extinction law is often an assumed quantity, its behavior across different sightlines, wavelength ranges, and total extinction is still an open question \citep[e.g.][]{Wang:2014xr, Nataf:2016dd, Schlafly:2016cr, Hosek:2018lr, Wang:2019jw, Nogueras-Lara:2019fc}.
Thus, the extinction law may be an important source of systematic error, and can be easily investigated 
with \texttt{SPISEA}.
Additionally, since full filter integration is used for the synthetic photometry, subtle effects such as the curvature in a reddening vector at high extinction \citep[e.g., due to the change in effective wavelength between two filters; ][]{Kim:2005jw} can be captured \citep[][]{Hosek:2018lr}. 

\begin{lstlisting}[language=Python, caption={Making a Theoretical Cluster Isochrone}, label=code:iso]
# Import necessary packages 
from spisea import synthetic, evolution
from spisea import atmospheres, reddening
import numpy as np

# Define isochrone parameters
logAge = np.log10(5*10**6.) # Age in log(years)
AKs = 0.8 # extinction in Ks-band mags
dist = 4000 # distance in parsec
metallicity = 0 # Metallicity in [M/H]

# Define evolution/atmosphere models and extinction law
evo_model = evolution.MISTv1() 
atm_func = atmospheres.get_merged_atmosphere
red_law = reddening.RedLawHosek18b()

# Specify filters for synthetic photometry. Here we 
# use  the HST WFC3-IR F127M, F139M, and F153M filters
filt_list = ['wfc3,ir,f127m', 'wfc3,ir,f139m', 
'wfc3,ir,f153m']

# Make Isochrone object. We will use the IsochronePhot 
# object since we want synthetic photometry. 
#
# Note that is calculation will take a few minutes to run, 
# unless this isochrone has been generated previously.
my_iso = synthetic.IsochronePhot(logAge, AKs, dist, 
                            metallicity=0,
                            evo_model=evo_model, 
                            atm_func=atm_func,
                            red_law=red_law, 
                            filters=filt_list)
                            
# Access the astropy table containing the individual 
# points in the isochrone. The columns described in 
# Table 1
iso_tab = my_iso.points
\end{lstlisting}

\begin{figure*}
\begin{center}
\includegraphics[scale=0.28]{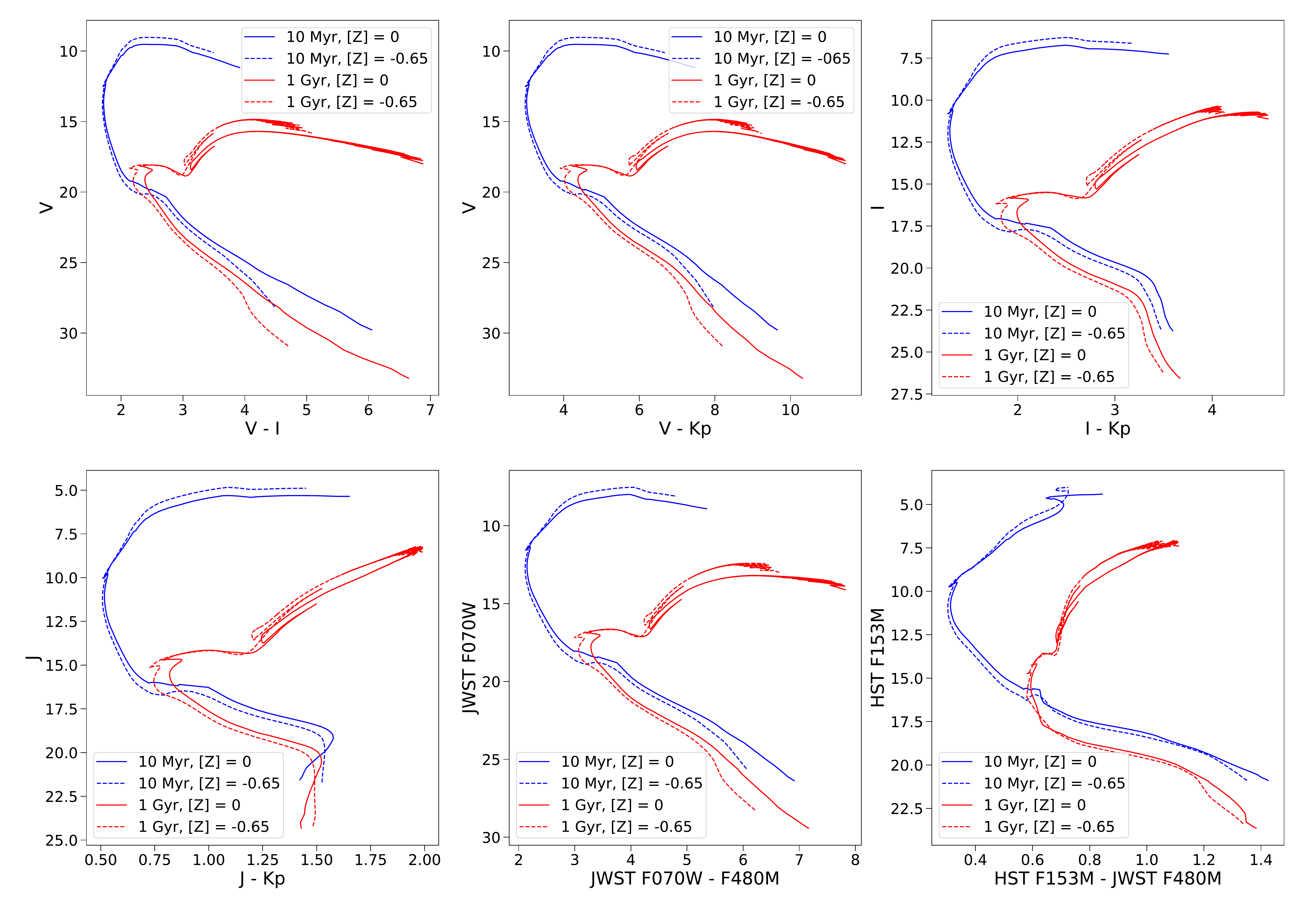}
\caption{CMDs of a young (10 Myr; blue) and old (1 Gyr; red) cluster isochrone across a range of filters at optical/near-infrared wavelengths. Two metallicities for each age are shown: [Z] = 0 (solar; solid line) and [Z] = -0.65 (dotted line). The isochrones have A$_{K}$ = 0.5 mag \citep[extinction law from ][]{Cardelli:1989qf} and a distance of 4000 pc. The filters represented are Johnson-Cousins V+I, Keck/NIRC2 J+Kp, HST F153M, and JWST F070W+F480M. The full set of pre-loaded filters can be found in Appendix \ref{app:models}, and more can be added by the user.}
\label{fig:iso_cmd}
\end{center}
\end{figure*}

The user can select what filters are used for synthetic photometry.
A suite of filters comes pre-loaded (Appendix \ref{app:models}), and new filters can be added by the user in a relatively straightforward way.
The pre-loaded filters span from $\sim$0.3 $\mu$m -- 5 $\mu$m and cover a range of telescopes/filter systems (Figure \ref{fig:iso_cmd}).

\subsection{\texttt{SPISEA} Clusters}
The code required to generate a synthetic cluster is shown in Listing \ref{code:clust}. 
Since \texttt{SPISEA} clusters requires an isochrone object as an input, the user has access to all of the customization options available to the isochrone object in addition to the ability to set the initial mass, IMF, multiplicity, differential extinction, and IFMR. 

\begin{lstlisting}[language=Python, caption={Making a Synthetic Cluster}, label=code:clust]
# Note: the code below assumes that the isochrone 
# has already been created as in Listing 1

# Import necessary packages 
from spisea import synthetic, ifmr
from spisea.imf import imf, multiplicity
import numpy as np

# Define stellar multiplicity properties. Here we 
# use the default multiplicity object.
# If no multiplicity desired, set this variable 
# to None
imf_multi = multiplicity.MultiplicityUnresolved()

# Define the IFMR. Here we use the default
# IFMR object. 
# If no IFMR is  desired, set this variable 
# to None
my_ifmr = ifmr.IFMR()

# Define the IMF. Here we'll use a broken
# power-law with the parameters from 
# Kroupa et al. (2001, MNRAS, 322, 231),
# and the multiplicity we defined.

# NOTE: when defining the power law slope 
# for each segment of the IMF, we define
# the entire exponent, including the negative sign. 
# For example, if dN/dm \propto m^-alpha,
# then you would use the value -2.3 to specify 
# an IMF with alpha = 2.3. 

massLimits = np.array([0.15, 0.5, 1, 120]) # mass segments
powers = np.array([-1.3, -2.3, -2.3]) # power-law exponents 
my_imf = imf.IMF_broken_powerlaw(massLimits, powers, 
imf_multi)

# Define the initial cluster mass
mass = 10**5 # Units: solar masses

# Make the cluster  object
cluster = synthetic.ResolvedCluster(my_iso, my_imf, mass, 
ifmr=my_ifmr)

# Access the astropy tables with the properties of 
# the star systems and the individual 
# companion stars. 
# The columns of the table are
# described in Table 2
star_systems = cluster.star_systems
companion_stars = cluster.companions

\end{lstlisting}

\begin{figure*}
\begin{center}
\includegraphics[scale=0.30]{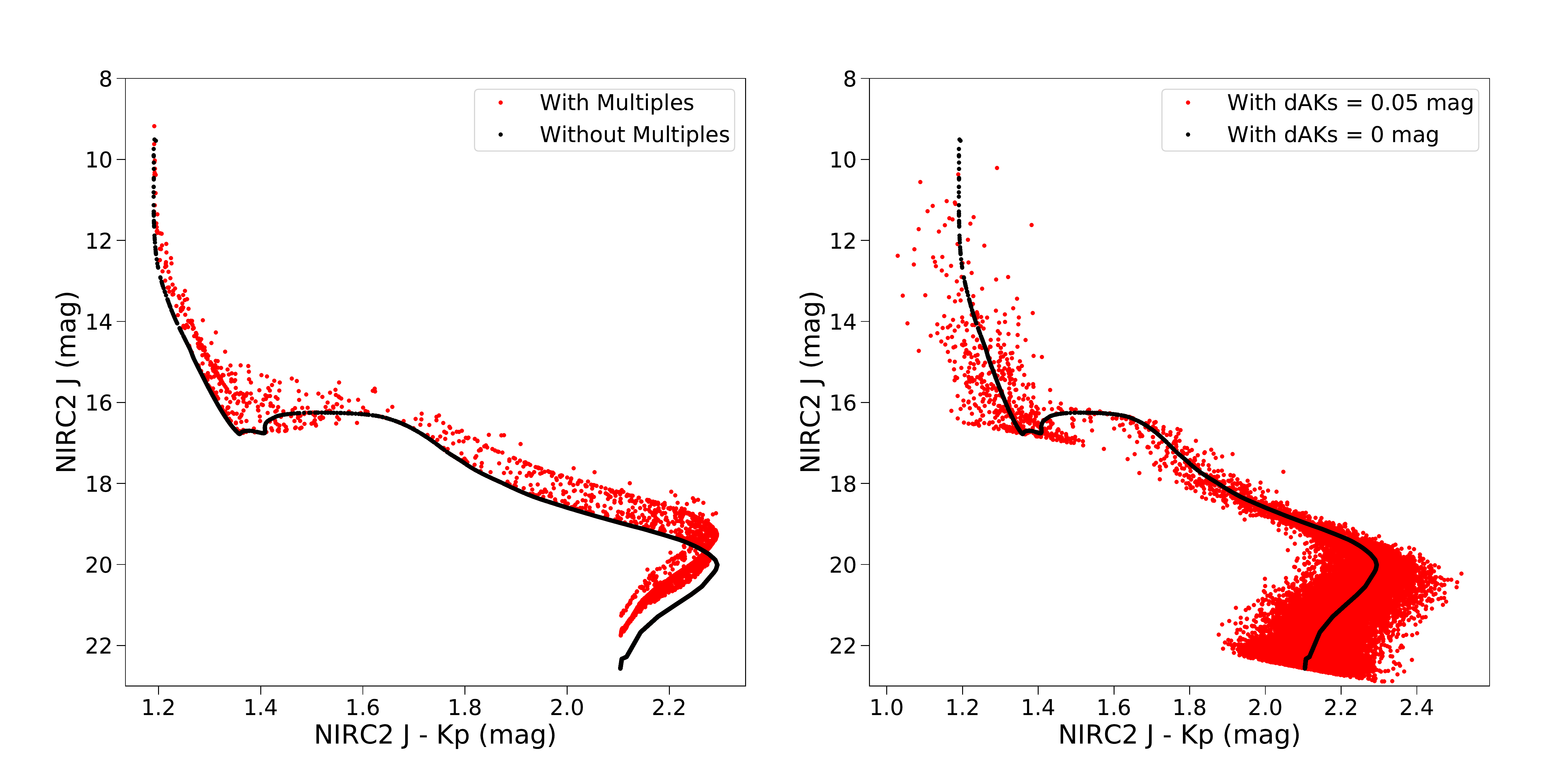}
\caption{\emph{Left:} A \texttt{SPISEA} cluster generated with (red) and without (black) the multiplicity as described by \citet{Lu:2013wo}. The presence of multiples generally makes star systems appear brighter and redder than single stars. \emph{Right:} A \texttt{SPISEA} cluster with differential extinction applied (dA$_{Ks}$ = 0.05 mag; red) versus one without (black). No multiple systems are included in order to isolate the impact of differential extinction on the CMD. For all isochrones, the \texttt{MISTv1} evolution models are used with the \texttt{get\_merged\_atmospheres} atmosphere models.}
\label{fig:cluster_cmd}
\end{center}
\end{figure*}

Figure \ref{fig:cluster_cmd} shows the impact that multiplicity and differential extinction has on the CMD of a star cluster. 
Both broaden the observed cluster sequence, albeit in different ways. 
The presence of unresolved multiples makes individual star systems appear brighter and/or redder than their single star counterparts, while differential extinction shifts stars both to the blue and red sides of the average cluster sequence. 
The unique way that multiplicity broadens the CMD can be used to statistically constrain the multiplicity properties of star clusters, though the impact of differential extinction, photometric errors, and stellar crowding must be considered \citep[e.g.][]{Hu:2010qp, de-Grijs:2013yi}. 

\begin{figure*}
\begin{center}
\includegraphics[scale=0.30]{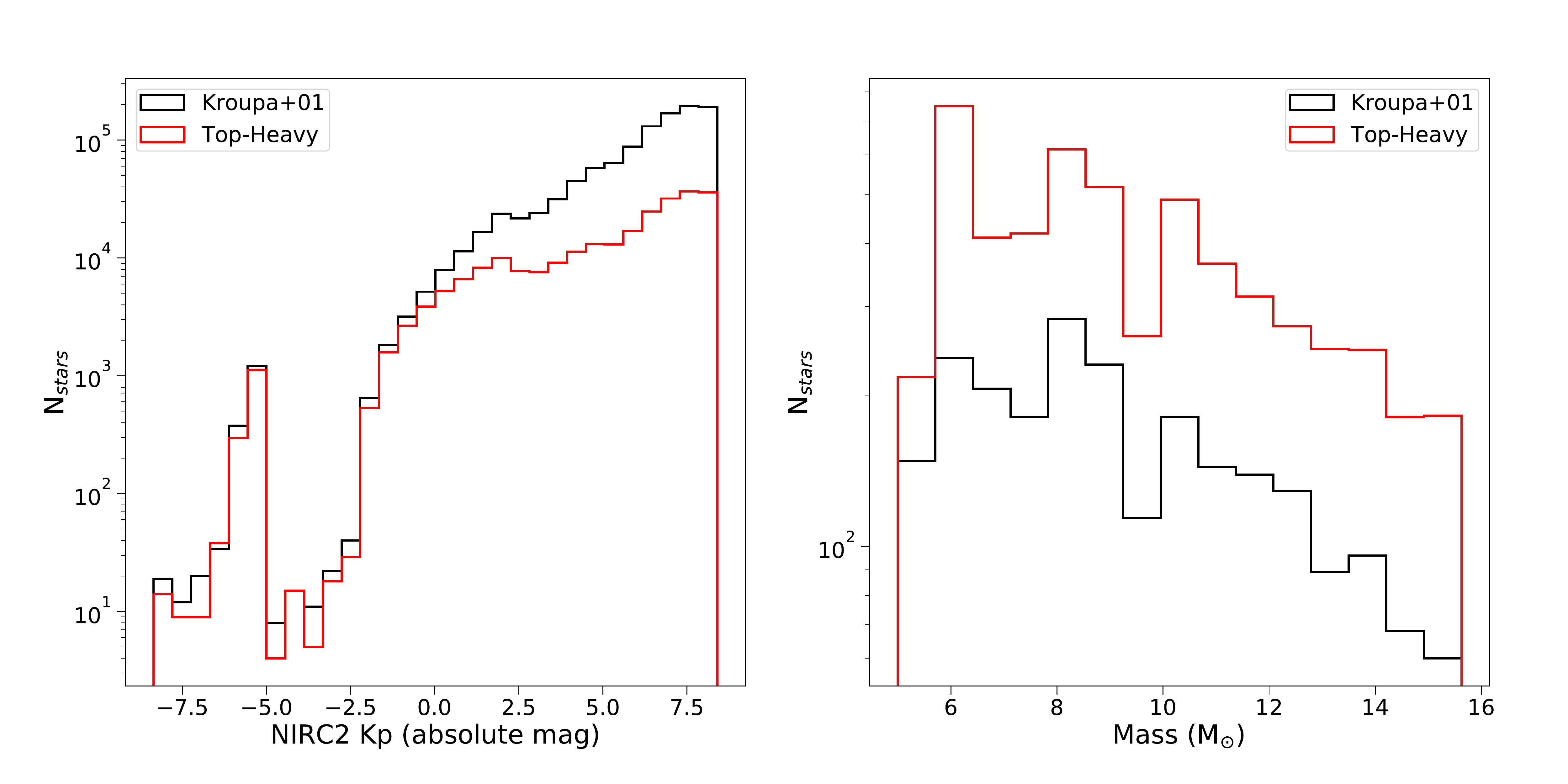}
\caption{\emph{Left:} NIRC2 Kp luminosity function of two clusters (10$^6$ M$_{\odot}$, 100 Myr, [Z] = 0), one from \citet[][black]{2001MNRAS.322..231K} and the other with the top-heavy IMF from \citet[][red]{Hosek:2019kk}. For a given total mass, the top-heavy IMF has a larger fraction of high-mass stars to low-mass stars relative to the Kroupa IMF. \emph{Right:} The black hole mass function for the two clusters, with the Kroupa IMF in black and top-heavy IMF in red. The top-heavy IMF produces more black holes than the Kroupa IMF, especially at higher masses.}
\label{fig:imf_comp}
\end{center}
\end{figure*}

The ability to generate clusters with different IMFs has made \texttt{SPISEA} a key component of IMF studies of the Young Nuclear Cluster \citep{Lu:2013wo} and the Arches Cluster \citep{Hosek:2019kk}. 
Figure \ref{fig:imf_comp} shows the Kp luminosity function for two identical clusters with different IMFs, one with a ``standard'' IMF of \citet{2001MNRAS.322..231K} and the other with a top-heavy IMF (e.g, a relative overabundance of high-mass stars) similar to \citet{Hosek:2019kk}.
With \texttt{SPISEA}, one can examine how the stellar population and compact remnants are expected to change with the IMF.
Figure \ref{fig:imf_comp} also includes the black hole mass function for both IMFs, as defined by the default IFMR object.

\begin{figure}
\begin{center}
\includegraphics[scale=0.50]{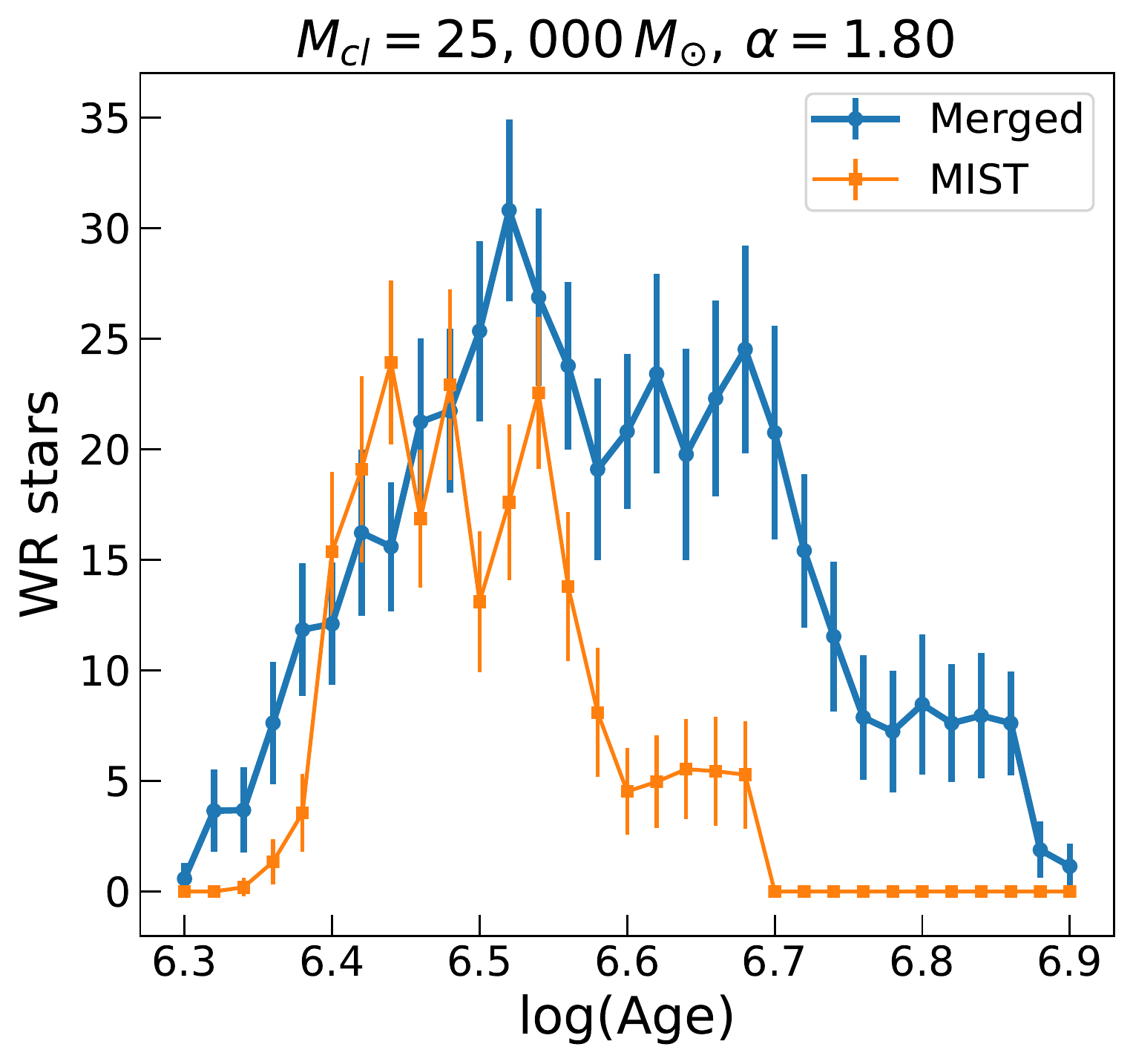}
\caption{Number of WR stars in synthetic clusters over different ages generated by \texttt{SPISEA} with \texttt{MergedBaraffePisaEkstromParsec} and \texttt{MISTv1} evolution models, given the cluster mass ($M_{cl} = 25,000 M_{\odot}$) and IMF slope ($\alpha$ = 1.80).}
\label{fig:WRstars}
\end{center}
\end{figure}

One can also assess the impact of different evolution and atmosphere models on a simulated star cluster by changing the isochrone that is used.
For example, the number ratio of massive Wolf-Rayet (WR) stars to other types of stars is a useful age indicator for a population \cite[e.g.][]{Meynet:1994dd}, though it can be affected by metallicity, stellar rotation, and binary star evolution \cite[e.g.][]{Ekstrom:2012qm, Georgy:2012uq, Dorn-Wallenstein:2018la}. 
\texttt{SPISEA} allows the user to examine how differences between evolution models impacts these predictions (Figure \ref{fig:WRstars}).
Note that \texttt{SPISEA} does not yet distinguish between different sub-classes of WR stars, such as WC and WN stars. 

\begin{figure}
\begin{center}
\includegraphics[scale=0.30]{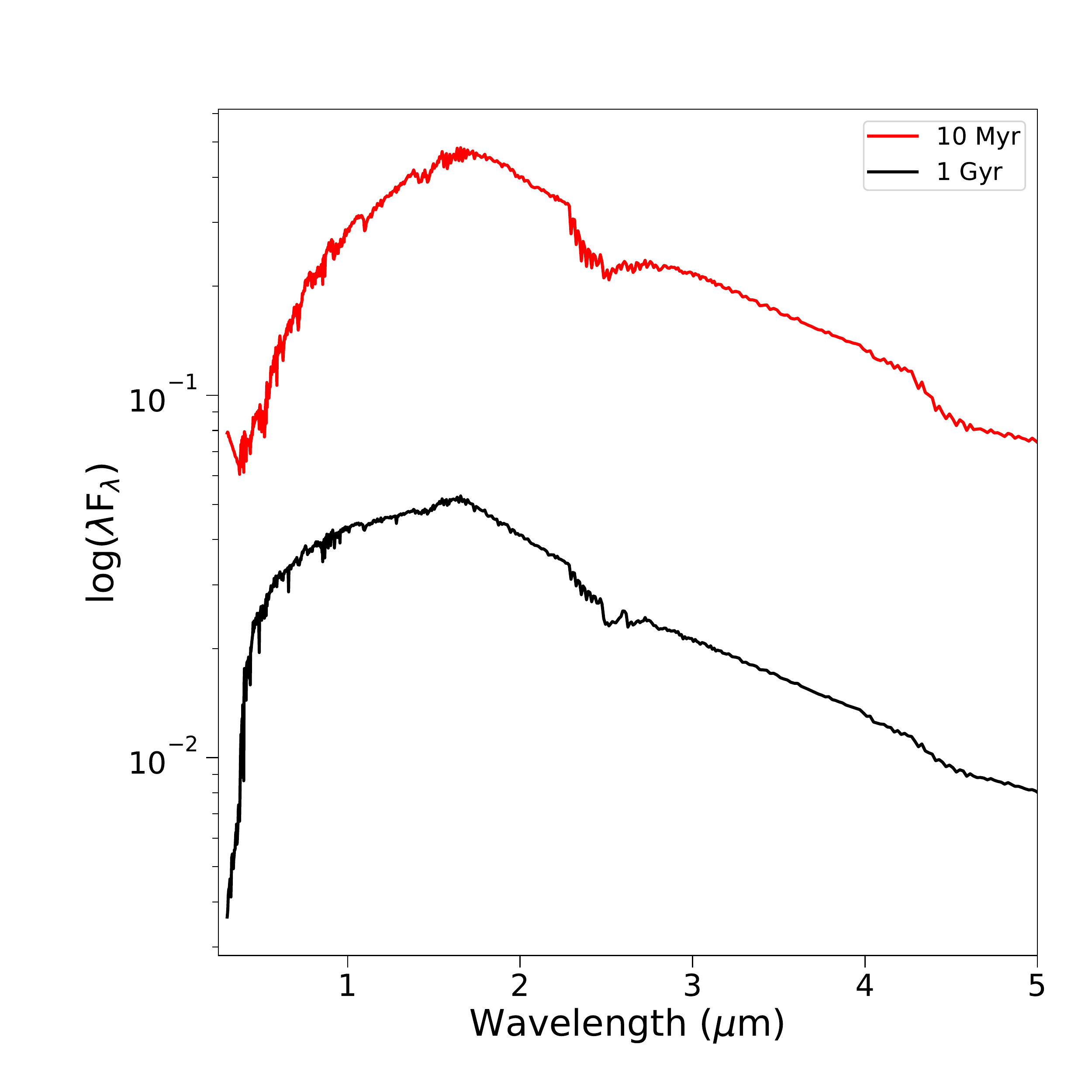}
\caption{Unresolved near-infrared spectra of a young (10 Myr, red) and old (1 Gyr, black) cluster. The young cluster is generated with an initial mass of 5x10$^3$ M$_{\odot}$, while the old cluster has an initial mass of 5x10$^4$ M$_{\odot}$. The differences in the composite spectra are driven by the differences in the stellar populations, in particular the presence of high-mass stars in the young population.}
\label{fig:clust_unres}
\end{center}
\end{figure}

In addition, \texttt{SPISEA} has the ability to sum the spectra of the individual stars to make a composite spectrum of the stellar population (Figure \ref{fig:clust_unres}). 
This can be used to analyze unresolved stellar populations. 
However, note that there is no treatment of nebular emission, which can have a significant impact on observations of very young clusters which remain enshrouded in leftover gas and dust leftover from formation \citep[e.g.][]{2009MNRAS.398..451M, Reines:2010zi}. 
In addition, \texttt{SPISEA} uses stellar atmospheres that assume local thermodynamic equilibrium (LTE), an assumption that breaks down for the most massive stars, which can dominate a composite spectrum.
Implementing non-LTE atmospheres is a priority in future code development ($\mathsection$\ref{sec:limit}), though the user can implement additional atmospheres of their choosing with the current release.

\subsection{Comparing to Observations}
\label{sec:praesepe}

\begin{figure}
\begin{center}
\includegraphics[scale=0.30]{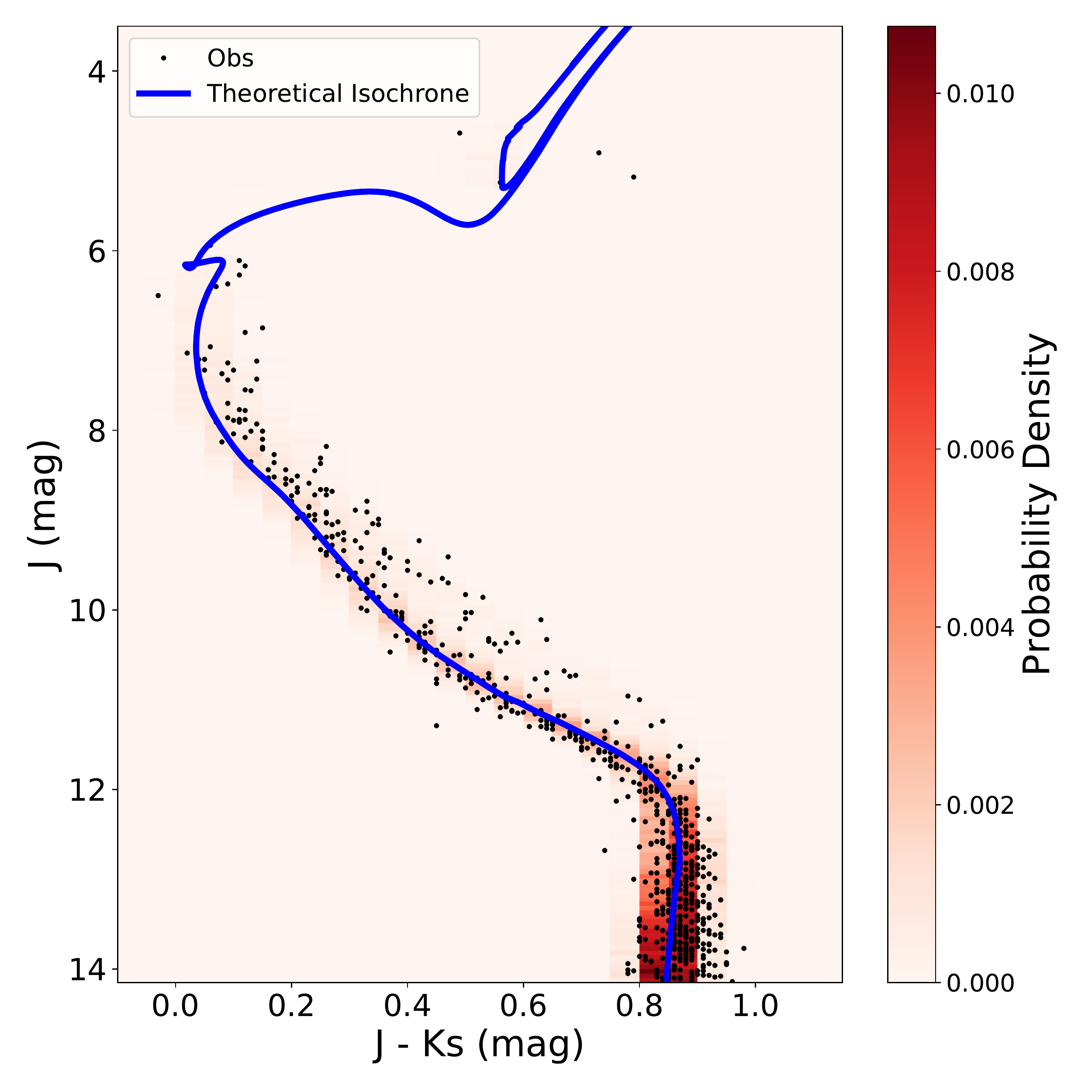}
\caption{A comparison of the observed 2MASS CMD of M44 \citep{Wang:2014sf} to a \texttt{SPISEA} cluster generated using the best-fit values from the literature. The observed cluster candidates are represented by the black points (typical uncertainty: 0.02 mags) and the theoretical isochrone by the blue line. The CMD is shaded according to the probability density distribution of the stars in the synthetic cluster. The \texttt{SPISEA} cluster provides a generally good match to the observed data.}
\label{fig:praesepe}
\end{center}
\end{figure}

\citet{Hosek:2019kk} use \texttt{SPISEA} to compare the observed CMD of the Arches Cluster to synthetic color-magnitude diagrams of model clusters with different input properties.
As an additional demonstration, Figure \ref{fig:praesepe} compares the observed 2MASS CMD of the Praesepe Cluster (M44) to a \texttt{SPISEA} cluster with the best-fit properties from the literature.
The observed data contain cluster candidates with M $\gtrsim$ 0.3 M$_{\odot}$ identified via kinematic and photometric properties by \citet{Wang:2014sf}.
We adopt the following properties for the synthetic cluster: age = 590 Myr \citep{Gossage:2018hh}, A$_{K}$ = 0 mag \citep[][measure A$_{v}$ = 0.08 mag, which is negligible at K-band]{Taylor:2006yw}, distance = 179 pc \citep{Gaspar:2009ds}, [Z] = 0 \citep[][obtain a slightly super-solar metallicity for M44, but solar metallicity is the closest in the grid of MIST models currently available in \texttt{SPISEA}]{Boesgaard:2013lr}, and a standard Kroupa IMF \citep{Boudreault:2012kw}.
We adopt the default \texttt{MultiplicityUnresolved} object ($\mathsection$\ref{sec:Mult}) for the multiplicity properties of the cluster, and use the MIST stellar evolution models and \texttt{get\_merged\_atmosphere} atmosphere models to generate the synthetic stars. 
We simulate photometric errors by perturbing the synthetic photometry of each star by a random amount drawn from the typical photometric uncertainty of the observations (0.02 mags).
We see that the \texttt{SPISEA} CMD is a generally good match to the observations, though a detailed analysis is beyond the scope of this paper.

\subsection{Code Performance}
\label{sec:runtime}
When simulating a cluster with \texttt{SPISEA}, the majority of the computation time is spent calculating synthetic photometry in the isochrone. 
On an 3 GHz Intel Xenon processor, the time required to compute an isochrone from scratch is $\sim$1 -- 5 mins, depending on the age of the cluster, the evolution/atmosphere models used and number of photometric filters chosen.
However, once the isochrone is generated, the process of building the cluster is fast: one can make a \texttt{ResolvedCluster} object with an initial mass of 10$^4$ M$_{\odot}$ in $\sim$1 s. 
Thus, after pre-generating a grid of isochrones, one can quickly build many clusters with different properties within likelihood functions for forward modeling purposes. 

Building an \texttt{UnresolvedCluster} object is more time intensive, since the process of assigning spectra to each star in the cluster is relatively slow. One can make an unresolved cluster with an initial mass of 10$^4$ M$_{\odot}$ cluster with a \citet{2001MNRAS.322..231K} IMF in $\sim$15 s.

\section{\texttt{SPISEA} Limitations and Future Development}
\label{sec:limit}
The first release of the \texttt{SPISEA} code has several limitations that present opportunities for future code development. 
Some of these limitations require relatively straightforward adjustments to the current code, and will be addressed in future code releases: 

\begin{itemize}[labelindent=1.5em,leftmargin=1em]
\item{\emph{Very Hot Star Models}: All of the stellar atmosphere models that are pre-loaded into \texttt{SPISEA} assume LTE, an assumption that breaks down for the most massive stars. This is especially true for Wolf-Rayet stars, which have extreme stellar winds that must be accounted for \citep[e.g.][]{Hillier:1998df, Grafener:2002yi}. While the user has the flexibility to add their own hot star atmospheres in the short term, we plan to implement them among the pre-loaded atmospheres in the long term.}
\item{\emph{Very Cool Star Models}: The pre-loaded \texttt{SPISEA} evolution models stretch to the brown-dwarf limit at 0.08 M$_{\odot}$ \citep[via the models from][]{Baraffe:2015yg}, but do not extend into the brown dwarf regime. In addition, the lowest temperature atmospheres only extend to 1200 K. Similar to the very hot star models, very cool star models can be implemented individually by the user, but ideally they would be included in the base package in the future.}
\item{\emph{Additional Metallicity Support}: Currently, the only non-solar stellar evolution models that comes with \texttt{SPISEA} is from MIST \citep{Choi:2016en}. However, other non-solar metallicity models exist, for example from the Geneva group for main sequence/post-main sequence stars \citep[e.g.][]{Georgy:2013lk} and the Pisa group for pre-main sequence stars \citep{Tognelli:2011fr}. These can be implemented by the user for now, and will be pre-loaded in future code releases.}
\item{\emph{Additional IMF Functional Forms}: \texttt{SPISEA} offers the user full control in defining an IMF with a broken power-law functional form ($\mathsection$\ref{sec:IMF}). However, a log-normal functional form has been proposed for M $\lesssim$ 1 M$_{\odot}$, with a power-law for M $\gtrsim$ 1 M$_{\odot}$ \citep[e.g.][]{Chabrier:2003wb}. While the true functional form of the IMF is not yet clear, improved observational facilities will allow for IMF measurements at the low masses necessary to distinguish between these two functional forms \citep[e.g.][]{El-Badry:2017qf, Hosek:2019uq}. 
Thus, implementation of the log-normal form of the IMF will be very useful for future analyses. This will be included in future code releases, but can be implemented by the user in the meantime.}
\item{\emph{Complex Star Formation Histories}: Currently, \texttt{SPISEA} only produces SSPs and is not equipped to produce a composite stellar population with a distribution of ages or metallicities. While most star clusters are assumed to be SSPs, recent work shows this assumption may not be valid for globular clusters \citep[e.g.][]{Piotto:2015pb}, and certainly more complex star formation histories are required to model galaxy-wide stellar populations. While it straightforward to approximate a composite population by combining a series of individual SSPs \citep[e.g.][]{2003MNRAS.344.1000B, Lam:2020vl}}, implementing a star formation history module within the code itself is an avenue for future work.
\end{itemize}

In addition, future releases of \texttt{SPISEA} will also have new and updated stellar evolution and atmosphere models, IFMRs, and multiplicity properties pre-loaded as they become available. 

Addressing other limitations will require more substantial code development:
\begin{itemize}[labelindent=1.5em,leftmargin=1em]
\item{\emph{Binary Star Evolution}: While \texttt{SPISEA} has a treatment for unresolved stellar multiplicity ($\mathsection$\ref{sec:Mult}), only single star stellar evolution models are available in the pre-loaded set. However, it has been found that interactions between binary stars can have a significant impact on stellar evolution, particularly for high-mass stars \citep[e.g.][]{Hurley:2002fk, Eldridge:2017rc, Dorn-Wallenstein:2018la}. \texttt{SPISEA} does not yet produce orbital properties of multiple systems (e.g. orbital periods and eccentricities), and so interactions cannot be modeled. In addition, modeling binary interactions is computationally intensive, and so a major expansion of the current code or an interface with an existing binary evolution codes would be required.}
 \item{ \emph{Theoretical Dust Extinction and Nebular Emission}: \texttt{SPISEA} does not include any radiative transfer or photoionization calculations, and so theoretical dust extinction curves and nebular emission cannot be calculated within the code. Instead, the user is required to do such calculations outside of \texttt{SPISEA} and implement their own extinction law objects as desired. Designing an interface between \texttt{SPISEA} and existing radiative transfer codes such as \texttt{cloudy} \citep{Ferland:2013kl} would be a significant undertaking, but can be explored if there is interest.}
\end{itemize}

\section{User Contributions}
\label{sec:contributions}
We encourage community input and contributions to \texttt{SPISEA} through the GitHub page given in $\mathsection$\ref{sec:intro}. 
Any bugs, questions, or feature requests should be reported via the issue tracker\footnote{https://github.com/astropy/SPISEA/issues}. 
If the user wishes to add features themselves, we ask that they fork or branch off of the current development repository, make their changes, 
and then submit merge and pull requests.
The contributions will be added (and attribution given) in future releases of the code.

\section{Conclusions}
\label{sec:conclusions}
We introduce \texttt{SPISEA}, an open-source \texttt{Python} code to generate SSPs. 
The modular interface of \texttt{SPISEA} offers unparalleled flexibility in defining the IMF, IFMR, extinction law, stellar multiplicity properties, and the stellar evolution and atmosphere model grids used to generate synthetic star clusters.
Example code outputs include cluster color-magnitude diagrams in a multitude of filters (also defined by the user), HR-diagrams, stellar mass functions, and compact object populations. 
\texttt{SPISEA} is available on \href{https://github.com/astropy/SPISEA}{GitHub} and \href{https://spisea.readthedocs.io/en/latest/}{ReadtheDocs}. We encourage input and contributions from the community.

\acknowledgements 
The authors thank the anonymous referee whose feedback increased the clarity of the paper, as well as the alpha testers whose feedback greatly improved the usability of \texttt{SPISEA}: Zhuo Chen, Devin Chu, Lauren Corlies, Shashank Dholakia, Michael Medford, Kerianne Pruett, and Gail Zasowski. The authors also thank BJ Fulton for a template for adding code into latex documents. M.W.H. and J.R.L. acknowledge support from NSF AAG (AST-1518273), and M.W.H. and S.J. acknowledge support from the Heising-Simons Foundation. This research has made use of NASA's Astrophysics Data System. 
\\
\\
\software{\texttt{pysynphot} \citep{STScI-Development-Team:2013fd}, \texttt{Astropy} \citep{Astropy-Collaboration:2013kx,Astropy-Collaboration:2018ws}}

\bibliographystyle{aasjournal}
\bibliography{main.bib}

\appendix

\section{Pre-Loaded SPISEA Options}
\label{app:models}
Here we describe the set of pre-loaded models in the initial release of \texttt{SPISEA}. 
The set of evolution and and atmosphere model grids is shown in Table \ref{tab:models}, extinction laws in Table \ref{tab:extinction}, and photometric filters in Table \ref{tab:filters}. 
Additional models can be added by the user.

\begin{deluxetable*}{l c c c c c c} 
\tablecaption{Evolution and Atmosphere Models}
\tabletypesize{\scriptsize}
\tablehead{
\multicolumn{7}{c}{\underline{Evolution Models}} \\
\colhead{Model Name} & \colhead{Mass Range} & \colhead{log(Age) Range} & \colhead{Metallicity Range} & \colhead{} & \colhead{} & \colhead{Ref} \\
\colhead{} & \colhead{M$_{\odot}$} & \colhead{Years} & \colhead{[Fe/H]} & \colhead{} & \colhead{} & \colhead{}
}
\startdata 
MIST v1.2 & 0.1 -- 300 & 6.0 -- 10.01 & -4.0 -- 0.5 &  &  & \citet{Choi:2016en} \\
MergedBaraffePisaEkstromParsec & 0.08 -- 120 & 6.0 -- 10.09  & 0 & &  & see Appendix \ref{app:merged} \\
Parsec & 0.1 -- 65 & 6.6 -- 10.12 & 0 & &  & \citet{Bressan:2012ya} \\
Baraffe15 & 0.07 -- 1.4 & 5.7 -- 10.0 & 0 & &  & \citet{Baraffe:2015yg} \\
Ekstrom12 & 0.8 -- 300 & 6.0 -- 8.0 & 0 & &  & \citet{Ekstrom:2012qm} \\
Pisa & 0.2 -- 7 & 6.0 -- 8.0 & 0 & &  & \citet{Tognelli:2011fr}\\
\hline
\multicolumn{7}{c}{\underline{Atmosphere Models}} \\
Model Name & T$_{eff}$ Range & log $g$ Range & Metallicity Range & $\lambda$ Range & Resolution\tablenotemark{a} & Ref \\
  &  K &  cgs & [Fe/H] & $\mu$m & $\lambda$ / $\Delta \lambda$ &  \\
\hline
get\_merged\_atmosphere & 3200 -- 50000 & \emph{b} & \emph{b} & \emph{b} & \emph{b} & see Appendix \ref{app:merged} \\
get\_castelli\_atmosphere & 3500 -- 50000  & 0 -- 5.0 & -2.5 -- 0.2 & 0.1 -- 10 & $\sim$250 & \citet{Castelli:2004yq} \\
get\_phoenixv16\_atmosphere & 2300 -- 12000 & 0.0 -- 6.0 & -4.0 -- +1.0 & 0.05 -- 5.5 & 100,000 -- 500,000 & \citet{Husser:2013ts} \\
get\_BTSettl\_2015\_atmosphere & 1200 -- 7000 & 2.5 -- 5.5 & 0 & 0.01 -- 30 & 2000 -- 700,000 & \citet{Baraffe:2015yg} \\
get\_BTSettl\_atmosphere\tablenotemark{d} &  2600 -- 7000 &  4.5 -- 5.5 &  -2.5 -- 0.5  &  0.1 -- 6.9 & 20,000 -- 250,000 & \citet{Allard:2012dq, Allard:2012zl} \\
get\_kurucz\_atmosphere & 3000 --- 50000 & 0 -- 5.0 & -5.0 -- 1.0 & 0.1 -- 10 & $\sim$250 & \emph{c} \\
get\_phoenix\_atmosphere & 2100 -- 69000 &  & -4.0 -- 0.5 & 0.001 -- 995 & $\sim$280 & \citet{Allard:2003eu, Allard:2007lq} \\
get\_wd\_atmosphere\tablenotemark{e} & --  & -- & -- & 0.1 -- 3.0 & 200 -- 500,000 & \citet{Koester:2010wd} \\
\enddata
\tablenotetext{a}{Spectral resolution of original atmosphere grid (often a function of $\lambda$, so approximate range reported here). As discussed in $\mathsection$\ref{sec:atmos}, the spectral resolution is degraded to R $\sim$ 250 by default for synthetic photometry. However, the user can choose to use the original high-resolution spectra if desired.}
\tablenotetext{b}{Depends on which model atmosphere grid is being used at user-specified temperature; see Appendix \ref{app:merged}}
\tablenotetext{c}{http://www.stsci.edu/hst/observatory/crds/k93models.html}
\tablenotetext{d}{T$_{eff}$ and log $g$ range depends on metallicity; reported values are the minimum range available. See documentation for exact ranges at each metallicity.}
\tablenotetext{e}{White Dwarfs only. If outside of \citep{Koester:2010wd} grid, will use blackbody spectrum instead}
\label{tab:models}
\end{deluxetable*}

\begin{deluxetable*}{l c c c c } 
\tablecaption{Extinction Laws}
\tabletypesize{\scriptsize}
\tablehead{
\colhead{Name} & \colhead{$\lambda_0$\tablenotemark{a}} & \colhead{$\lambda$ Range} & \colhead{User Inputs} & \colhead{Ref} \\
\colhead{} & \colhead{$\mu$m} & \colhead{$\mu$m} & \colhead{} 
}
\startdata 
RedLawReikeLebofsky & 2.12 & 0.365 - 13.0 &  & \citet{Rieke:1985dw} \\
RedLawCardelli & 2.174 & 0.3 -- 3.0 & R(V) & \citet{Cardelli:1989qf}\\
RedLawRomanZuniga07 & 2.134 & 1.24 -- 7.76  & &  \citet{Roman-Zuniga:2007yq} \\
RedLawFitzpatrick09 & 2.14 & 0.7 -- 3.0 & $\alpha$, R(V) & \citet{Fitzpatrick:2009ys} \\
RedLawNishiyama09 & 2.14 & 0.5 -- 8.0 &  & \citet{Nishiyama:2008wa, Nishiyama:2009fc} \\
RedLawFritz11 & 2.14 & 1.0 -- 19.0 &  & \citet{Fritz:2011cr} \\
RedLawDamineli16 & 2.159 & 0.44 -- 4.48 &  & \citet{Damineli:2016no} \\
RedLawSchlafly16 & 2.14 & 0.5 -- 4.8 & A$_H$ / A$_{Ks}$, x & \citet{Schlafly:2016cr} \\
RedLawHosek18\tablenotemark{b} & 2.14 & 0.7 -- 3.545 &  & \citet{Hosek:2018lr} \\
RedLawHosek18b & 2.14 & 0.7 -- 3.545 &  & \citet{Hosek:2019kk} \\
RedLawNoguerasLara18 & 2.15 & 0.8 -- 2.8 &  & \citet{Nogueras-Lara:2018uq}\\
RedLawPowerLaw & ---  & --- & $\alpha$, $\lambda_0$ & ---\\
\enddata
\tablenotetext{a}{Wavelength defined such that A$_{\lambda_0}$ / A$_{Ks}$ = 1}
\tablenotetext{b}{RedLawHosek18 is depreciated and the user should instead use RedLawHosek18b.}
\label{tab:extinction}
\end{deluxetable*} 

\begin{deluxetable*}{l c c } 
\tablecaption{Photometric Filters}
\tabletypesize{\scriptsize}
\tablehead{
\colhead{Telescope/System} & \colhead{Filters} & \colhead{Ref} \\
}
\startdata 
2MASS & J, H, K$_s$ & \citet{Cohen:2003ay} \\
CTIO/OSIRIS & H, K & 1 \\
DeCam & u, g, r, i, z, Y & \citet{Abbott:2018fm} \\
Gaia & G, Gbp, Grp & \citet{Evans:2018rc}\\
Hubble Space Telescope & see \texttt{pysynphot} documentation & -- \\
Johnson-Cousins & U, B, V, R, I & \citet{Johnson:1966jx} \\
Johnson-Glass & J, H, K & \citet{Bessell:1988im} \\
James Webb Space Telescope/NIRCAM & see website & 2 \\
Keck/NIRC & H, K & 3 \\
Keck/NIRC2 & J, H, Hcont, K, Kp, Ks, Kcont,  & \\
  & Lp, Ms, Brgamma, FeII & 4 \\
NACO & J, H, K & 5 \\
PanStarrs 1& g, r, i, z, y & \citet{Tonry:2012oz} \\
UKIRT & J, H, K &  \citet{Hewett:2006zt} \\
VISTA & Z, Y, J, H, K & 6 \\
\enddata
\tablenotetext{1}{http://www.ctio.noao.edu/soar/content/ohio-state-infrared-imagerspectrograph-osiris}
\tablenotetext{2}{https://jwst-docs.stsci.edu/display/JTI/NIRCam+Filters\#NIRCamFilters-filt\_trans}
\tablenotetext{3}{https://www2.keck.hawaii.edu/inst/nirc/}
\tablenotetext{4}{https://www2.keck.hawaii.edu/inst/nirc2/filters.html}
\tablenotetext{5}{https://www.eso.org/sci/facilities/paranal/instruments/naco/inst/filters.html}
\tablenotetext{6}{http://casu.ast.cam.ac.uk/surveys-projects/vista/technical/filter-set}
\label{tab:filters}
\end{deluxetable*} 

\section{Description of Merged Evolution and Atmosphere Models}
\label{app:merged}
\texttt{SPISEA} comes with one set of merged stellar evolution models and one set of merged stellar atmosphere models. 
These are created in order to take advantage of the strengths of different model sets in different parameter spaces, such as different stellar masses, temperatures, population ages, evolutionary stage, etc. 
These merged grids are currently only available at solar metallicities.

The merged stellar evolution object is \texttt{evolution.MergedBaraffePisaEkstromParsec}.
It is comprised of 4 recent stellar evolution models: Baraffe15 \citep{Baraffe:2015yg}, Pisa \citep{Tognelli:2011fr}, Ekstrom/Geneva \cite[both with and without rotation;][]{Ekstrom:2012qm}, and Parsec v1.2 \citep{Bressan:2012ya}.
Which models are used depends on the population age.
If logAge $<$ 7.4, then Baraffe15 is used for 0.08 M$_{\odot}$ -- 0.4 M$_{\odot}$, Pisa is used from 0.5 M$_{\odot}$ to the highest mass available (typically between 5 -- 7 M$_{\odot}$), and the Ekstrom/Geneva models are used from the highest mass in the Pisa models to 120 M$_{\odot}$.  
In the transition region between 0.4 M$_{\odot}$ -- 0.5 M${\odot}$, a linear interpolation between the Baraffe15 and Pisa models is used.
If logAge $>$ 7.4, Parsec v1.2 is used for the full mass range.  

This merged method was chosen to emphasize the strengths of the different models. 
For example, the Ekstrom/Geneva grid offers coverage of young main sequence and post-main sequence stars, but does not include the pre-main sequence.
Meanwhile, the combination of the Pisa and Baraffe15 grids offer coverage of young pre-main sequence stars down to the hydrogen burning limit. 
Additionally, Parsec is well suited for old star populations, where the Ekstrom/Geneva grid only covers ages up to 100 Myr.

The merged stellar atmosphere grid is defined by the \texttt{atmospheres.get\_merged\_atmospheres} object. 
It contains a mix of ATLAS9 \citep{Castelli:2004yq}, PHOENIX v16 \citep{Husser:2013ts}, BTSettl \citep{Baraffe:2015yg}, and Koester10 \citep{Koester:2010wd} atmospheres. 
It is described in $\mathsection$2.2 of \citet{Lam:2020vl}. 
Briefly, which atmosphere grid is used depends on the stellar temperature and evolutionary state.
For stars, the ATLAS9 grid is used for stars with T$_{eff}$ $>$ 5500 K, the PHOENIX grid is used for 5000 K $<$ T$_{eff}$ $<$ 3800 K, and the BTSettl grid is used for 3200 K $<$ T$_{eff}$ $<$ 1200 K. 
For temperatures at transition temperatures between grids (e.g. 5000 K -- 5500 K), an average atmosphere between the two model grids is used.
For white dwarfs with known physical properties (i.e., those included in the MISTv1 evolution models), the Koester10 atmospheres are used. 
If the white dwarf properties lie outside the Koester10 model grid, then a blackbody curve is used instead.

\end{document}